\documentclass[12pt]{iopart}
\usepackage{graphicx}

\newcommand {\bx} {\mbox{\boldmath $x$}}

\newcommand{\calI}{{\cal I}}

\begin{document}

\title[Statistical properties of entropy production]{Statistical properties of entropy production 
	derived from fluctuation theorems}

\author{Neri Merhav$^{(1)}$ and Yariv Kafri$^{(2)}$}

\address{(1) Department of Electrical Engineering, Technion, Haifa 32000, Israel. 
(2) Department of Physics, Technion, Haifa 32000, Israel}

\begin{abstract}
Several implications of well--known fluctuation theorems, on the
statistical properties of the entropy production, are 
studied using various approaches. We begin by deriving 
a tight lower bound on the variance of 
the entropy production for a given mean of this random variable.
It is shown that the Evans--Searles fluctuation theorem alone imposes a significant
lower bound on the variance only 
when the mean entropy production is very small. It
is then nonetheless demonstrated that upon
incorporating additional information 
concerning the entropy production, this lower bound can be significantly
improved, so as to capture extensivity properties. 
Another important aspect of the fluctuation properties of the
entropy production is the
relationship between the mean and the variance, on the one hand, and
the probability of the event where the entropy production is negative, on the
other hand. Accordingly, we derive 
upper and lower bounds on this probability in terms
of the mean and the variance.
These bounds are tighter than previous bounds that can be 
found in the literature. Moreover, they are tight in the sense that there
exist probability distributions, satisfying the Evans--Searles fluctuation
theorem, that achieve them with equality.
Finally, we present a general method for generating a
wide class of inequalities that must be
satisfied by the entropy production. We use this method to derive several new
inequalities which go beyond the standard derivation of the second law.
\end{abstract}

\maketitle

\section{Introduction}

It has been recently realized that time 
reversal symmetry implies that some
exact results, termed {\it fluctuation theorems}, hold for systems which are driven arbitrarily
far from thermal equilibrium \cite{review}. 

As an example, 
which will be of particular interest in this paper, 
consider an isolated system in a certain equilibrium state, which is driven to another
state by varying an external control parameter $\lambda$ over a time interval $0\leq t \leq T$, 
according to a certain protocol $\{\lambda(t),~0\le t\le T\}$. 
Following the application of the protocol, a difference 
$\sigma$ is developed between the entropy at the final state 
and the entropy at the initial state of the system. This difference is also
called the {\it entropy production.}
One can now imagine the same system being driven using a time--reversed 
protocol $\tilde{\lambda}(t)=\lambda(T-t)$. When the time--reversed protocol 
obeys $\tilde{\lambda}(t)=\lambda(t)$, 
it can be shown that time reversal symmetry 
implies the Evans--Searles fluctuation theorem \cite{EvansSearles2002,Der1}
\begin{equation}
\frac{p( \sigma)}{p(- \sigma)} = e^{\sigma} \;,
\label{EvSe}
\end{equation}
where $p(\sigma)$ is the probability density function of the entropy
production $\sigma$.
This immediately yields
\begin{equation}
\left< e^{- \sigma}\right> = 1,
\label{entropyrel}
\end{equation}
where as usual, we assume that the entropy change in the system 
is much smaller than its total entropy.
As is well known, the second law 
of thermodynamics, $\left<\sigma \right>\ge 0$, is easily obtained by
applying Jensen's inequality, $\left<e^{-\sigma}\right>\ge e^{-\left<\sigma
\right>}$, to the left--hand side of eq.\ (\ref{entropyrel}).
Other, directly related 
examples of fluctuation theorems include 
the Jarzynski equality, the Crooks theorem, 
and the Gallavotti--Cohen relation 
\cite{Evans_Cohen_Gallavotti,Gallavotti_Cohen,JarzynskiPRL2004,
JarzynskiPRL1997,CrooksPRE1999,JarzynskiPRE1997,JarzynskiJStatMech2004,CrooksJStatPhys1998}. 

Beyond their elegance and their pure academic value, some
of the fluctuation theorems have been suggested as possible tools 
for measuring {\it equilibrium} 
quantities from averages over repeated experiments of a {\it non--equilibrium}
process \cite{review}, which is a considerably interesting idea.
For instance, the Jarzynski equality paves the way to
measure free--energy differences 
from non--equilibrium measurements of work 
performed on (or by) the system. However, as is
evident from eq.\ (\ref{entropyrel}), events 
with a negative entropy production 
play an important role, which makes such a measurement impossible for
macroscopic systems.

As mentioned above, when Jensen's inequality is applied to eq.\
(\ref{entropyrel}), the second law is immediately obtained. Moreover, since 
the exponential function is strictly convex, Jensen's inequality becomes
an equality if and only if $\sigma$ is a degenerate random variable, which
takes on the value $\sigma=0$ with probability one. This corresponds to a
perfectly reversible process where no entropy is produced. In any other case,
where Jensen's inequality is a strict one, $\sigma$ becomes a
non--degenerate random variable, and hence it fluctuates about its mean,
which then
must be strictly positive. Thus, eq.\ (\ref{entropyrel}) already
tells us an important qualitative fact: 
Irreversibility, strictly positive mean entropy
production, and fluctuations (e.g., in terms of positive variance) appear 
always together. 
However, eq.\ (\ref{entropyrel}) 
does not immediately tell us much in the quantitative level. In particular,
when applying Jensen's inequality, it is felt that a great 
deal of valuable information, concerning the statistics
of the entropy production, is lost. 

It is the purpose of this paper to explore some quantitative restrictions
on the probability distribution of the entropy production 
that arise from the fluctuation theorems. Throughout most of the paper, 
we consider systems which obey the Evans--Searles fluctuation theorem, i.e.,
systems that satisfy eq.\ (\ref{EvSe}). 
Some restrictions are rather obvious. Consider for example 
the mean entropy production 
\begin{equation}
m=\langle \sigma \rangle 
\end{equation}
and the variance 
\begin{equation}
v=\langle \sigma^2 \rangle - \langle \sigma \rangle^2. 
\end{equation}
In the small mean entropy production limit, 
one can expand eq.\ (\ref{entropyrel}) to leading 
order and obtain the simple relation $v=2m$. 
The same restriction holds, of course, also in the Gaussian case (even without
taking the small mean entropy
limit), as only the first two
cumulants are non--zero in this case.

In this paper, we derive the most stringent restrictions, that can possibly be
posed by the fluctuation theorems, on the shape of the distribution. 
In particular, we first derive a lower bound on the variance $v$
of the entropy production $\sigma$ 
as a function of the mean entropy production $m$. 
The bound we derive is tight in the 
sense that there is a probability distribution 
of the entropy production, which achieves the bound with equality. 
It turns out, from this bound, that the 
Evans--Searles fluctuation theorem alone imposes a
meaningful lower bound on the variance $v$ only when the mean entropy production
$m$ is very small.
It is nonetheless demonstrated that upon
incorporating additional statistical (and/or physical) information
concerning the entropy production, this lower bound can be significantly
improved, so as to capture extensivity properties, which mean that in the
thermodynamic limit, it is plausible that 
both $m$ and $v$ should scale linearly with time and with the
system size, and so, they should be proportional to each other.

Next, we derive upper and lower bounds on the probability of 
negative entropy production, $\mbox{Pr}\{\sigma \le 0\}$,
as a function of the mean entropy production $m$ and the variance $v$. 
To the best of our knowledge, these bounds are better than related bounds,
previously derived using Jensen's inequality \cite{Jar2008}. They are
again tight in the sense that there exist probability distributions,
satisfying the Evans--Searles fluctuation theorem, which achieve them
with equality. The interesting fact, in this context, is that
we are actually obtaining a bound on the large deviations rate function
of the probability of the rare event  
$\{\sigma \le 0\}$. As will be shown later, if $\sigma$ is an extensive random
variable, then this probability can decay no faster than $e^{-m}$, which is
exponential in the system size (or time).
The upper and lower bounds are also {\it universal} (in the
probability distribution of $\sigma$) and the 
important fact is that they provide non--trivial
assessments on $\mbox{Pr}\{\sigma \le 0\}$, a quantity 
which is not easily measurable by
experiments, in terms of $m$ and $v$, 
which are measurable in principle. 

Our last result is about
a general analysis tool to be applied to eq.\ (\ref{entropyrel})
in order to derive a wide class of inequalities 
that involve the entropy production. These inequalities are, in general, more 
powerful than the standard Jensen inequality, and some of them lead to
certain variations of the second law, $\langle\sigma\rangle\ge 0$.
For example, among other things, we prove that 
the probability of the event $\{\sigma\le \alpha\}$, for any deterministic
parameter $\alpha$, is bounded from above according to
\begin{equation}
\mbox{Pr}\{\sigma\le \alpha\}\le
\exp\left\{\langle \sigma \rangle|_{\sigma \leq \alpha}\right\}, 
\end{equation}
where the average is conditional on $\sigma \leq \alpha$.
Thus, the second law is obtained as a special case, with $\alpha\to\infty$.
Another example of an inequality from this class is
\begin{equation}
\langle \sigma e^{-\sigma} \rangle \leq 0, 
\end{equation}
which tells us that although
$\sigma$ is non--negative on the average, as the second law asserts, 
the event of negative entropy production still has enough probabilistic weight
so as to make the average of $\sigma$, weighted by the function $e^{-\sigma}$
(which favors negative values of $\sigma$), negative rather than positive.

The remaining part of the
paper is structured as follows: In Sec.\ 2, we derive lower bounds 
on the variance of the entropy production given the mean. In Sec.\ 3, 
lower and upper bounds are derived for the probability of rare events 
with a negative entropy production. In Sec.\ 4, we present the method 
for deriving inequalities as mentioned above. Finally, we conclude in
Sec.\ 5.

\section{A lower bound on the variance of the entropy production}

Consider a system which obeys eq.\ (\ref{EvSe}), with $\left<\sigma\right>=m$,
where $m$ is given.
First, observe that since $p(-\sigma)=e^{-\sigma}p(\sigma)$, we have 
\begin{equation}
\int_{-\infty}^{+\infty} \mbox{d}\sigma \cdot p(\sigma)= 
\int_0^{+\infty} \mbox{d}\sigma p(\sigma)(1+e^{-\sigma})=1,
\end{equation}
which means that instead of considering a real valued random variable $\sigma$, 
taking both positive and
negative values, 
and distributed according to $p(\sigma)$,
we can consider, equivalently, a positive random variable,
distributed according to 
\begin{equation}
q(\sigma)=p(\sigma)(1+e^{-\sigma}),~~~\sigma\ge 0.
\end{equation}
The mean $m$ can then be expressed, in terms of $q$, according to
\begin{eqnarray}
m &=& \int_{-\infty}^{+\infty}\mbox{d}\sigma\cdot 
\sigma p(\sigma)\\
&=&\int_0^{+\infty}\mbox{d}\sigma\cdot
p(\sigma)\sigma(1-e^{-\sigma})\\
&=&\int_0^{+\infty}\mbox{d}\sigma\cdot 
q(\sigma)\sigma\cdot \frac{1-e^{-\sigma}}{1+e^{-\sigma}}\\
&=&\int_0^{+\infty}\mbox{d}\sigma\cdot 
q(\sigma)\sigma\tanh\left(\frac{\sigma}{2}\right)\\
&=&\int_0^{+\infty}\mbox{d}\sigma\cdot 
q(\sigma)f(\sigma),
\end{eqnarray}
where we have defined
\begin{equation}
f(\sigma)\equiv
\sigma \tanh\left(\frac{\sigma}{2}\right).
\end{equation}
Similarly, the second moment is
\begin{equation}
\left<\sigma^2\right> =\int_0^{+\infty}
\mbox{d}\sigma\cdot\sigma^2p(\sigma)(1+e^{-\sigma})
=\int_0^{+\infty}\mbox{d}\sigma\cdot
\sigma^2q(\sigma).
\end{equation}
In simple words, we have used the Evans--Searles fluctuation theorem
in order to transform 
a two--sided random variable $\sigma$, governed
by $p(\sigma)$, into a one--sided random variable ($\sigma \ge 0$),
whose probability density function is given by
$q(\sigma)=p(\sigma)(1+e^{-\sigma})$ and we henceforth denote expectations
under $p$ and under $q$, by $\left<\cdot\right>_p$ and
$\left<\cdot\right>_q$, respectively.
Thus, the 
constraint $\left< \sigma \right>_p=m$,
in the domain of the original two--sided random variable, is equivalent to
the constraint $\left<f(\sigma)\right>_q=m$ 
in the domain of the transformed, one--sided random variable.

\begin{figure}[h!t!b!]
\centering
\includegraphics[width=8.5cm, height=8.5cm]{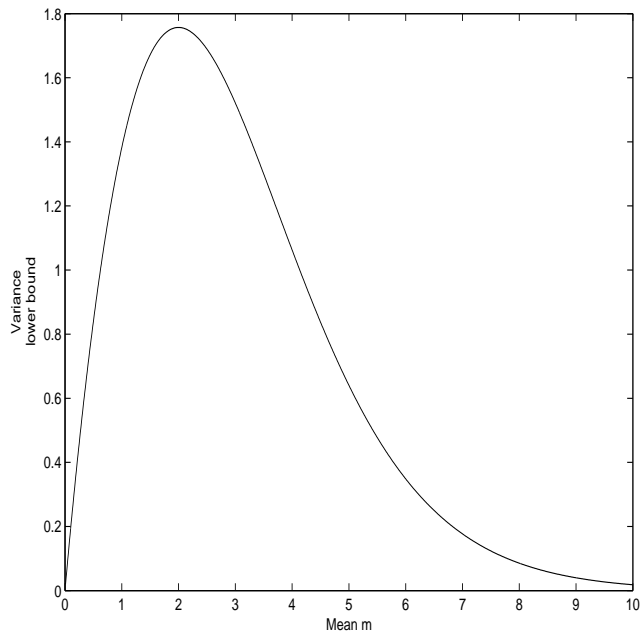}
\caption{A plot of the lower bound on the variance, $h^2(m)-m^2$, as a
function of $m$.}
\label{gmu1}
\end{figure}

Since $f$ is a monotonically strictly increasing function for $\sigma\ge 0$,
it is clearly invertible in this range, 
and we shall denote the inverse function of $f$
by $h$. I.e., $\mu=f(\sigma)$ if and only if $\sigma=h(\mu)$. We next
observe that $h^2(\mu)=[h(\mu)]^2$, 
which is obviously the inverse function of $f(\sqrt{\sigma})$,
is a convex function.\footnote{This follows from the fact that
$f(\sqrt{\sigma})$ is
monotonically increasing and concave for $\sigma\ge 0$, as can easily be
checked from the derivatives of this function.} Thus, we readily obtain
the following lower bound on the second moment in terms of $m$:
\begin{eqnarray}
\left<\sigma^2\right>_p&=&
\left<\sigma^2\right>_q\\
&=&\left<h^2(f(\sigma))\right>_q\\
&\ge & 
h^2(\left<f(\sigma)\right>_q)\\
&=& h^2(m),
\end{eqnarray}
where the inequality follows from the application of Jensen's inequality
to the convex function $h^2$.
Equality is obtained when $\sigma$ is deterministic under $q$, i.e.,
\begin{equation}
q(\sigma)\equiv q^*(\sigma)=\delta(\sigma-h(m)), 
\end{equation}
or equivalently,
\begin{equation}
\label{twodeltas}
p(\sigma)=p^*(\sigma)\equiv\frac{1}{1+e^{-h(m)}}\cdot\delta(\sigma-h(m))+
\frac{e^{-h(m)}}{1+e^{-h(m)}}\cdot\delta(\sigma+h(m)).
\end{equation}
Using the above result, we see 
that the variance of the entropy production can be bounded by
\begin{equation}
v \geq h^2(m) - m^2.
\label{bound1}
\end{equation}
This is the central result of this section. The function $h^2(m)-m^2$ is
depicted in Fig.\ \ref{gmu1}.

We comment that the above analysis holds, 
not only for the second moment. It can be 
generalized straightforwardly from the second moment,
$\left<\sigma^2\right>$, to every higher moment of the form
$\left<|\sigma|^k\right>$, where $k$ is any real number larger than 2
(i.e., $k$ does not have to be an integer).
The more general result is then
\begin{equation}
\left<|\sigma|^k\right>\ge h^k(m)\equiv [f^{-1}(m)]^k,
\end{equation}
where equality is universally achieved by the same density function $p^*$
as before.

Returning to the second moment, it is easy to see that 
$h(x) \geq x$ (with equality only at $x=0$), 
so the lower bound on the variance is positive.
For small $m$, $h^2(m)\approx 2m$, and so, to this order,
$v \geq 2m$, in agreement with the simple relation 
obtained by expanding eq.\ (\ref{entropyrel}). 
For large values of $m$, we have to leading 
order $h(m)=m(1+2e^{-m/2})$ and the bound 
decays as $4e^{-m/2}$. This behavior is easy to understand since
the minimum variance distribution, for a
given $m$, is achieved by a pair of delta functions, as seen above in
eq.\ (\ref{twodeltas}).
For large values of $m$, the contribution of the Dirac 
function at the negative value of $\sigma$ becomes essentially
irrelevant to the variance, due to the exponential weighting
of the Evans--Searles theorem. As advertised 
in the Introduction, 
in this context, the fluctuation theorems provide
useful information on the variance of the entropy production 
distribution only for relatively small values of the mean entropy production. 

The reason for this behavior is simple: The Evans--Searles fluctuation
theorem merely relates the probability density $p$ at negative
values of $\sigma$ to those at the corresponding positive values, but
as $m$ grows without bound (the
thermodynamic limit), most of the probability mass goes for positive
values of $\sigma$ anyway, and so, the information concerning negative
values becomes essentially irrelevant. Therefore, it is understood that
the Evans--Searles theorem {\it alone} cannot possibly give useful
information about the thermodynamic limit, and this is not because of
a possible weakness in the derivation of the lower bound (which is tight,
in the sense of being achieved by $p^*$).
This means that in order to obtain
more meaningful bounds, one must incorporate additional information.

To demonstrate this fact, suppose that the additional information we have
is given by the extra constraint
\begin{equation}
\mbox{Pr}\{|\sigma|\le r\}=b,
\end{equation}
where $r>0$ and $b\in[0,1]$ 
are given\footnote{In practice, this additional information
may be a result of an experimental measurement.} and where
we will assume that $f(r) < m$. How does the lower bound on the variance $v$
change in the presence of this additional constraint?

Once again, we refer to 
the one--sided distribution $q$. The two constraints 
now read $\left<f(\sigma)\right>_q=m$ and $q\{\sigma\le r\}=b$. 
Next, we note that every probability measure 
$q$, satisfying the two constraints on the positive reals, can be represented
as a mixture of two probability density functions, $q_1$ and $q_2$, as follows:
\begin{equation}
q(\sigma)=bq_1(\sigma)+(1-b)q_2(\sigma),
\end{equation}
where the support of $q_1$ is (a subset of) the interval $[0,r]$ and the
the support of $q_2$ is (a subset of) $(r,\infty]$. We next denote
expectations with respect to $q_1$ and $q_2$ by $\left<\cdot\right>_1$ and 
$\left<\cdot\right>_2$, respectively. 
Let us define
$m_1=\left<f(\sigma)\right>_1$ and 
$m_2=\left<f(\sigma)\right>_2$, where, 
of course, $b m_1+(1-b) m_2=m$, $m_1\le f(r)$. 
Now, from the same considerations as in the proof of the first 
bound derived above, we have 
\begin{equation}
\left<\sigma^2\right>_1\ge
h^2(m_1) 
\end{equation}
and
\begin{equation}
\left<\sigma^2\right>_2\ge
h^2(m_2). 
\end{equation}
This implies
\begin{eqnarray}
	\left<\sigma^2\right>&=& b
	\left<\sigma^2\right>_1+(1-b)
	\left<\sigma^2\right>_2 \nonumber \\
	&\ge& bh^2(m_1)+(1-b)h^2(m_2) \nonumber\\
	&=& bh^2(m_1)+(1-b)h^2\left(\frac{m-bm_1}{1-b}\right)\nonumber\\
	&\ge& bh^2(f(r))+(1-b)h^2\left(\frac{m-bf(r)}{1-b}\right)\nonumber\\
	&=& br^2+(1-b)h^2\left(\frac{m-bf(r)}{1-b}\right) \;,
\end{eqnarray}
where the second inequality follows from the limitation $m_1\le f(r)$
and the convexity of $h^2$, which implies that the minimum of
$bh^2(m_1)+(1-b)h^2\left(\frac{m-bm_1}{1-b}\right)$, subject
to the constraint $m_1\le f(r)$, is achieved for $m_1=f(r)$ (provided
that $f(r) < m$, as assumed). The above lower bound 
is achieved as both inequalities in the last chain become equalities, and
this is the case for
\begin{equation}
q(\sigma)=b\delta(\sigma-r)+(1-b)\delta\left(\sigma- 
h\left(\frac{m-bf(r)}{1-b}\right)\right),
\end{equation}
which means two pairs of Dirac delta functions in the domain of the
original, one--sided random variable, governed by $p$.

In view of the foregoing discussion
on the first lower bound, it is now interesting to examine what 
happens in the limit of large mean entropy production $m$. 
In this regime, it makes sense to let $r$ increase with $m$,
so as to keep $b$ approximately constant and thus to avoid a situation
where the event $|\sigma|\le r$ becomes a rare one. 
In the lower bound, when both $r$ and $m-bf(r)$ are large,
$f(r)\approx r$ and 
\begin{equation}
h\left(\frac{m-bf(r)}{1-b}\right)\approx 
\frac{m-bf(r)}{1-b}\approx 
\frac{m-br}{1-b}
\end{equation}
and so,
\begin{eqnarray}
\left<\sigma^2\right>&\geq& br^2+(1-b)\left(\frac{m-br}{1-b}\right)^2\nonumber\\
&=&\frac{br^2+m^2-2m rb}{1-b}\\
&=&m^2+ \frac{b(r-m)^2}{1-b},
\end{eqnarray}
which means that the asymptotic lower bound on the variance is
\begin{equation}
\label{approxlb}
v \geq  \frac{b(r-m)^2}{1-b}.
\end{equation}
This is especially suitable in situations where 
central limit theorem arguments hold, 
and then, in the vicinity of the peak, the distribution can be approximated 
by a Gaussian with a mean proportional to the 
variance (both being extensive variables). Then
$b$ remains approximately constant if the 
dependence of $r$ upon $m$ is chosen to be 
$r=r(m)\equiv m-c\sqrt{m}$, where $c$ is some arbitrary positive constant. 
In this case, the lower bound (\ref{approxlb}) becomes
\begin{equation}
v \geq  \frac{bc^2m}{1-b},
\end{equation}
which is indeed extensive (proportional to $m$), as may be expected.
This is very different from the earlier lower bound, which vanishes as $m$
tends to infinity.

\section{Bounds on the probability of negative entropy production}

Eq.\ (\ref{EvSe}) embeds a certain symmetry property
of the entropy production about the origin. In this section, 
we employ this symmetry to derive upper and lower bounds 
on the probability of the (rare) event that the entropy decreases. 
This serves as another measure of fluctuations 
in the entropy production. Namely, we derive bounds on 
$\mbox{Pr}\{\sigma\le 0\} = \int_{-\infty}^0d\sigma p(\sigma)$, which
depend merely on the mean $m$ and the variance $v$.\footnote{More precisely,
our lower bound will depend merely on $m$, whereas the upper bound will
depend on both $m$ and $v$.}

For the lower bound, we use the simple inequality
\begin{equation}
e^{-x}\ge e^{-\alpha}-e^{-\alpha}(x-\alpha),
\end{equation}
which follows from the simple fact that the exponential function
is lower bounded by the affine function tangential to it at any
point $(\alpha,e^{-\alpha})$ on the curve, where $\alpha$ is an arbitrary real
parameter, whose value is left for our choice. Using this inequality, we now
have
\begin{eqnarray}
\mbox{Pr}\{\sigma \le 0\}
&=&\int_{-\infty}^0\mbox{d}\sigma\cdot p(\sigma) \nonumber\\
&=&\int_0^{\infty}\mbox{d}\sigma e^{-\sigma}\cdot p(\sigma)\\
&\ge& \int_0^{\infty}\mbox{d}\sigma[e^{-\alpha}-e^{-\alpha}(\sigma-\alpha)]\cdot
p(\sigma)\\
&=&(1+\alpha)e^{-\alpha}\mbox{Pr}\{\sigma>0\}-e^{-\alpha}\int_0^\infty\mbox{d}\sigma\cdot
\sigma p(\sigma)\\
&=&(1+\alpha)e^{-\alpha}[1-\mbox{Pr}\{\sigma \le 0\}]-e^{-\alpha}\int_0^\infty\mbox{d}\sigma \cdot
\sigma p(\sigma),
\end{eqnarray}
and so
\begin{equation}
\mbox{Pr}\{\sigma\le 0\}\ge
\frac{(1+\alpha-m_+)e^{-\alpha}}
{1+(1+\alpha)e^{-\alpha}}=
\frac{1+\alpha-m_+}
{e^{\alpha}+\alpha+1},
\end{equation}
where we have denoted 
$m_+=\int_0^\infty\mbox{d}\sigma\cdot \sigma p(\sigma)$.
Since this holds for any real $\alpha$, then
\begin{equation}
\mbox{Pr}\{\sigma\le 0\}\ge
\sup_{\alpha}\frac{1+\alpha-m_+}
{e^{\alpha}+\alpha+1}.
\end{equation}
The maximum is achieved for $\alpha=\psi(m_+)$, 
where $\psi(\cdot)$ is the inverse of the function
$\phi(\alpha)=\alpha/(1+e^{-\alpha})$.
This yields
\begin{equation}
\mbox{Pr}\{\sigma\le 0\}\ge 
\frac{1+\psi(m_+)-m_+}{e^{\psi(m_+)}+\psi(m_+)+1}.
\end{equation}
In the the thermodynamic limit, $m_+\approx m$ and $\alpha\approx m_+$,
and so,
\begin{equation}
\label{lower}
\mbox{Pr}\{\sigma\le 0\}\ge \frac{1}{e^{m}+m+1}\sim
e^{-m}.
\end{equation}
We observe then that in the thermodynamic limit,
if $m$ is an extensive variable,
the probability of a negative entropy 
production cannot decay exponentially faster than $e^{-m}$.
While this lower bound is universal (in the sense of being independent
of the actual probability distribution of $\sigma$), it is nevertheless
a tight bound in the sense that it is achieved by a certain distribution
that satisfies the Evans--Searles fluctuation theorem (again, given by a
pair of Dirac delta functions).

Next we derive an upper bound on the 
probability of negative entropy production. 
To this end, we use the following inequality
which applies for any $x\ge 0$ and $a\ge 0$:
\begin{equation}
e^{-x}\le e^{-a}-e^{-a}(x-a)+g(x-a)^2
\end{equation}
where
\begin{equation}
g\equiv\frac{1-(a+1)e^{-a}}{a^2} \;.
\end{equation}
This upper bounds the exponential function $e^{-x}$ by a quadratic
function, tangential to the exponential function
at the point $x=a$, where the coefficient $g$
of the quadratic term is chosen so as to keep the quadratic function
above the exponential function for every positive $x$.
Denoting $s_+=\int_0^\infty\mbox{d}\sigma\cdot \sigma^2p(\sigma)$,
we have
\begin{eqnarray}
\mbox{Pr}\{\sigma\le 0\}
&=&\int_0^{\infty}\mbox{d}\sigma e^{-\sigma}\cdot p(\sigma)\\
&\le& \int_0^{\infty}\mbox{d}\sigma\left[
e^{-a}-e^{-a}(\sigma-a)+g(\sigma^2-2a\sigma+a^2)
\right]\cdot p(\sigma)\\
&=& [(a+1)e^{-a}+a^2g]
\mbox{Pr}\{\sigma> 0\}-e^{-a}m_+ +g(s_+-2a m_+)\\
&=&\mbox{Pr}\{\sigma> 0\}-e^{-a}m_+ +g(s_+-2a m_+)\\
&=&1-\mbox{Pr}\{\sigma\le 0\}-e^{-a}m_+ +g(s_+-2a m_+)
\end{eqnarray}
and so,
\begin{equation}
\mbox{Pr}\{\sigma\le 0\}\le \frac{1+gs_+-2agm_+-e^{-a}m_+}{2}.
\end{equation}
Since this holds for every $a\ge 0$, 
\begin{eqnarray}
\mbox{Pr}\{\sigma\le 0\}&\le&\inf_{a\ge 0}\frac{1+gs_+-2ag m_+-e^{-a}m_+}{2}\\
&=&\frac{1}{2}+\frac{1}{2}\inf_{a\ge 0}
(gs_+-2ag m_+-e^{-a}m_+).
\end{eqnarray}
A minimization over $a$ yields $a=s_+/m_+$ so that,
\begin{eqnarray}
gs_+-2agm_+-e^{-a}m_+&=&gs_+-2gs_+-e^{-a}m_+\\
&=&-(gs_++e^{-a}m_+)\\
&=&-(gs_++m_+e^{-s_+/m_+})
\end{eqnarray}
where
$$g=\frac{m_+^2}{s_+^2}-\left(\frac{m_+}{s_+}+
\frac{m_+^2}{s_+^2}\right)e^{-s_+/m_+}.$$
Thus,
\begin{equation}
gs_++m_+e^{-s_+/m_+}=
\frac{m_+^2}{s_+}\left(1-e^{-s_+/m_+}\right)
\end{equation}
and so,
\begin{equation}
\mbox{Pr}\{\sigma\le 0\}\le
\frac{1}{2}-\frac{m_+^2}{2s_+}+\frac{m_+^2}{2s_+}e^{-s_+/m_+}.
\end{equation}
Similarly as the lower bound, the 
upper bound too is universal and tight in the sense defined above.
Here, in the thermodynamic limit, $m_+\to m$ and
$s_+\to s\equiv v+m^2$, which yields
\begin{eqnarray}
\mbox{Pr}\{\sigma\le 0\}&\le&
\frac{1}{2}\left(1-\frac{m^2}{s}\right)+\frac{m^2}{2s}e^{-s/m}\\
&=&\frac{v}{2s}+\frac{m^2}{2s}e^{-s/m}.
\end{eqnarray}
Using the bound obtained in the previous 
section (see eq.\ (\ref{bound1}) and 
the discussion following it), one may wonder what is the range of values that
this bound can possibly take for a given values of $m$ and $v$. One can check
that the minimum of the bound, in the large $m$ limit, 
is given when the variance is minimum, i.e., $v=4e^{-m/2}$,
so that to leading order, it is $2e^{-m/2}/m^2$. Its maximum value is 
achieved when $v \gg m^2$, and it behaves like $v/(2s)=1/(2m^2/v+2)$.
For comparison, we note that the well--known Chebychev inequality (that
also bounds the probability of interest in terms of $m$ and $v$)
yields
\begin{eqnarray}
\mbox{Pr}\{\sigma\le 0\}&=&
\mbox{Pr}\{m-\sigma\ge m\}\\
&\le&\frac{\left<(m-\sigma)^2\right>}{m^2}\\
&=&\frac{v}{m^2},
\end{eqnarray}
which is a weaker upper bound, that does not make use of the
fluctuation theorem.

\section{Inequalities based on Jensen's inequality with a change of measure}

As mentioned already in the Introduction, 
it is common practice to go from eq.\ (\ref{entropyrel})
to the second law via Jensen's inequality. 
As was also said in the Introduction, a great deal of information is
lost by this application of Jensen's inequality, while it is felt that
eq.\ (\ref{entropyrel}) has much more to tell.

In this section, 
we present a method by which 
a variety of stronger and more informative 
inequalities can be generated from eq.\ 
(\ref{entropyrel}) by applying Jensen's inequality in a somewhat more
sophisticated way, which allows a change of the probability measure,
where the new measure is subjected to optimization. Unlike the usual 
use of Jensen's inequality, here no information is lost at all since eq.\ 
(\ref{entropyrel}) can be `recovered' as special case.

First, let us recall the elementary fact that the 
entropy production $\sigma$ is actually a function of the random path (or the
trajectory) $\bx=\{x_t:~0\le t\le T\}$ taken by the system in phase space
as time runs from $t=0$ to $t=T$, where $x_t$ is the microscopic state at
time $t$. In other words, $\sigma$ should actually be denoted by
$\sigma(\bx)$. For example, in the case of Markovian
dynamics of a system controlled by an agent, $\sigma(\bx)=\beta
W_d(\bx)=\beta[W(\bx)-\Delta F]$, is the entropy production due to dissipated
work $W_d(\bx)$ at fixed temperature $1/\beta$, where $W(\bx)$ is the total work
and $\Delta F$ is the free--energy difference. 
This is then Jarzynski's equality. 

Let the probability law, that governs the trajectory $\bx$, be denoted by $P$.
Let $Q$ be an arbitrary alternative probability measure for
$\bx$. To avoid ambiguities, 
we denote, until further notice, expectations with respect to \ $P$ and
$Q$ by $\left<\cdot\right>_P$ and
$\left<\cdot\right>_Q$, respectively.
Consider now the following chain of inequalities, where before applying
Jensen's inequality, we change the underlying probability measure from $P$ 
to $Q$, and thereby obtain an expression that depends on $Q$:
\begin{eqnarray}
1&=&\left< e^{-\sigma(\bx)}\right>_P \nonumber\\
&=&\int\mbox{d}\bx P(\bx)e^{-\sigma(\bx)}  \nonumber\\
&=&\int\mbox{d}\bx Q(\bx)e^{-\sigma(\bx)+\ln[P(\bx)/Q(\bx)]}  \nonumber\\
&=&\left<e^{-\sigma(\bx)+\ln[P(\bx)/Q(\bx)]}\right>_Q  \nonumber\\
&\ge& e^{-\left<\sigma\right>_Q-D(Q\|P)}.
\end{eqnarray}
Here we have introduced the relative entropy 
$D(Q\|P)=\langle \ln \left[ Q({\bx})/P({\bx}) \right] \rangle_Q$. 
This is equivalent to the inequality
\begin{equation}
\label{basic}
\left<\sigma\right>_Q+D(Q\|P)\ge 0
\end{equation}
which holds true
for every probability measure $Q$. For $Q=P$, we are, of course, back to the
second law.
However, a stronger inequality is obtained upon minimizing the
left hand side with respect to $Q$ 
across some set of probability measures that includes
$Q=P$. The global minimum of the 
left hand side among {\it all} probability measures
(which is attained by $Q^*(\bx)=P(\bx)e^{-\sigma(\bx)}$)
turns out to be zero,
and so, this leads to an uninteresting
trivial identity. However, if the left 
hand side is minimized across some smaller set
of measures (e.g., those that maintain moments of certain functions or
physical quantities),
then one still obtains a non--trivial inequality, which is stronger
than the ordinary second law.

Another useful way to 
present inequality (\ref{basic}) is the following: Consider an
arbitrary non--negative path function $\Psi=\Psi(\bx)$ with
$\left<\Psi\right>_P\in (0,\infty)$ and let us select
\begin{equation}
Q(\bx)=\frac{P(\bx)\Psi(\bx)}{\int\mbox{d}\bx' P(\bx')\Psi(\bx')}=
\frac{P(\bx)\Psi(\bx)}{\left<\Psi\right>_P}.
\end{equation}
Then, obviously,
\begin{equation}
\left<\sigma\right>_Q=
\frac{\left<\Psi\cdot\sigma\right>_P}{\left<\Psi\right>_P},
\end{equation}
and
\begin{eqnarray}
D(Q\|P)&=&\int\mbox{d}\bx \cdot Q(\bx)\log\frac{Q(\bx)}{P(\bx)} \nonumber\\
&=&\int\mbox{d}\bx\cdot\frac{P(\bx)
\Psi(\bx)}{\left<\Psi\right>_P}\cdot
\log\frac{\Psi(\bx)}{\left<\Psi\right>_P}\nonumber\\
&=&\frac{1}{\left<\Psi\right>_P}\left(\left<\Psi\log
\Psi\right>_P-\left<\Psi\right>_P\log\left<\Psi\right>_P\right).
\end{eqnarray}
On substituting these expressions back into (\ref{basic}), we obtain
the following inequality, which is the main result of this section:
\begin{equation}
\left<\Psi\sigma\right>\ge \left<\Psi\right>\cdot\log\left<\Psi\right>-
\left<\Psi\log\Psi\right>,
\label{ineq}
\end{equation}
where we have omitted the subscript $P$ from the expectation operator since
now, all expectations are taken again with respect to the original measure $P$. 
This extends the inequality $\left<\sigma\right> \ge 0$ (the
second law) by correlating $\sigma(\bx)$ with an arbitrary
random variable $\Psi(\bx)$, measurable on the path. 
In other words, we now have a family of bounds
on `projections' of $\sigma(\bx)$ in the `directions' of all non--negative path
functions $\Psi(\bx)$, and not just in the one direction pertaining to
$\Psi(\bx)\equiv 1$, as in the second law.
The right hand side of this inequality
depends solely on the statistics of $\Psi$, and it is never positive due to
the convexity of the function $f(t)=t\log t$ for $t\ge 0$. 

The above lower bound is tight in the sense that there is a choice of
$\Psi(\bx)$ for which the inequality becomes an equality, and this
is $\Psi(\bx)=e^{-\sigma(\bx)}$. Since the left hand side of the inequality is
$\left<\sigma e^{-\sigma}\right>$, for this choice of $\Psi$, and it is
identical to the right hand side which is non--positive. It follows that
\begin{equation}
\left<\sigma e^{-\sigma}\right> \le 0.
\end{equation}
Note that the last inequality can also be derived 
directly from the non-positivity 
of the right hand side of eq.\ (\ref{ineq}). 
This means (similar to the usual interpretation 
of eq.\ (\ref{entropyrel})) that although 
$\sigma$ is non--negative on the average, as the
second law asserts, there is enough probabilistic weight to 
paths $\{\bx\}$
for which $\sigma < 0$, so that the last inequality must hold.

A slightly more general case arises 
for $\Psi=e^{-\alpha \sigma}$ ($\alpha$ being 
an arbitrary real parameter), which yields
\begin{equation}
\left<\sigma e^{-\alpha\sigma}\right> \ge \left<e^{-\alpha\sigma}\right>
\log\left<e^{-\alpha\sigma}\right>+\alpha\left<
\sigma e^{-\alpha\sigma}\right>\;,
\end{equation}
or equivalently,
\begin{eqnarray}
\left< \sigma e^{-\alpha \sigma}\right> 
&\ge& \frac{\left<e^{-\alpha\sigma}\right>
\log\left<e^{-\alpha \sigma}\right>}{1-\alpha}~~~~~\alpha < 1 \nonumber\\
\left<\sigma e^{-\alpha \sigma}\right> &\le&
\frac{\left< e^{-\alpha\sigma} \right>
\log \left< e^{-\alpha\sigma}\right>}{1-\alpha}~~~~~\alpha > 1\;.
\end{eqnarray}
Once again, the ordinary second law is obtained by substituting $\alpha=0$ in
the first of these two inequalities, which pertains to the case $\alpha < 1$.

Yet another interesting choice is $\Psi(\bx)=\calI\{\bx:~\sigma(\bx)\le
\alpha\}$, where $\calI\{\cdot\}$ is the indicator function of the event in
the braces and $\alpha$ is an arbitrary real parameter. 
In this case, our inequality tells us something about the
conditional expectation of $\sigma$ given that $\sigma(\bx)\le\alpha$:
\begin{equation}
\left<\sigma\right>\bigg|_{\sigma\le\alpha}\ge 
\log \mbox{Pr}\{\sigma \le\alpha\},
\end{equation}
where the ordinary second law is recovered as 
$\alpha\to\infty$.
It may be interesting to look at the last inequality as an upper bounded on
$\mbox{Pr}\{\sigma\le\alpha\}$:
\begin{equation}
\mbox{Pr}\{\sigma\le\alpha\}\le \exp\left\{\left<\sigma\right>|_{\sigma\le\alpha}\right\},
\end{equation}
which is interesting when $\left<\sigma
\right>|_{\sigma\le\alpha} < 0$. For $\alpha < 0$, this is certainly
the case. Note also that this is tighter than a straightforward use of the
Chernoff bound \cite{Jar2008,JarBB}, according to
\begin{equation}
\mbox{Pr}\{\sigma\le\alpha\}\le \left<e^{\alpha-\sigma}\right> = e^\alpha
\end{equation}
because $\left<\sigma\right>|_{\sigma\le\alpha}$ cannot exceed $\alpha$.

\section{Conclusion} 

In this paper, we have derived a series of bounds related to 
properties of the entropy production that stem from 
fluctuation theorems. These bounds illustrate that fluctuation theorems 
restrict the form of the probability distribution in a significant manner 
only for small mean entropy changes, but it is possible to
improve these bounds upon incorporating additional information. Furthermore, we derived rigorous 
lower and upper bounds on the probability of the rare event where the entropy 
production is negative. The bounds depend on the mean entropy 
production and its variance, two quantities which are more 
readily accessible to measurement than the probability of this rare event.
Finally, we presented a systematic way to derive inequalities which result 
from fluctuation theorems. These go beyond the standard derivation of the second law.

Acknowledgments: We are grateful for useful comments and discussions with Chris Jarzynski and Richard Blythe.

\end{document}